\documentclass[journal=jacsat,manuscript=article]{achemso}

\usepackage[version=4]{mhchem}
\usepackage{graphicx}
\usepackage{amsmath}
\setkeys{acs}{doi = true}
\usepackage{hyperref}
\DeclareUnicodeCharacter{2212}{\textminus}

\author{Jalen Macatangay}
\author{Alejandro Strachan}
\email{strachan@purdue.edu}
\affiliation{School of Materials Engineering and Birck Nanotechnology Center, Purdue University, West Lafayette, Indiana, 47907, USA}

\title{Nuclear Quantum Effects in Multi-Step Condensed Matter Chemistry: A Path Integral Molecular Dynamics Study of Thermal Decomposition}

\begin{document}

% Abstract
%%%%%%%%%%%%%%%%%%%%%%%%%%%%%%%%%
\begin{abstract}
    Nuclear quantum effects (NQEs) are often central to a predictive understanding of chemical reactions and rates. While their incorporation in gas-phase reactions is well established, studies involving condensed matter often neglect or approximate such effects. To clarify the role of NQEs in multi-step, multi-molecular reactions in a molecular crystal, we compare atomistic simulations of the thermal decomposition of the energetic material TATB using path integral molecular dynamics (PIMD), the more approximate quantum thermal bath (QTB), and classical MD (ClMD). PIMD samples the quantum canonical distribution by representing each atom as a string of beads (replicas), while QTB uses a frequency-dependent thermostat to reproduce the Bose-Einstein distribution. We find that PIMD results in faster chemical decomposition of the TATB crystal compared to ClMD, as the initial steps involve hydrogen transfer processes. Interestingly, some of the subsequent reactions (e.g. the formation of \ce{N2}) occur on identical timescales. The PIMD simulations also predict a reduction in overall activation energy by $\sim$8\% as compared to the classical result. As observed in model systems and simple unimolecular gas-phase reactions, the QTB significantly overestimates quantum acceleration of chemical reactions and the reduction in activation energy. A comparison of the kinetic energy operator in PIMD and the centroid dynamics provides insight into the physics behind the differences between the QTB and PIMD results.
\end{abstract}

% Introduction
%%%%%%%%%%%%%%%%%%%%%%%%%%%%%%%%%
\newpage
\section{Introduction}

% Introduction paragraph
The classical treatment of nuclei in molecular dynamics (MD) has enabled large-scale and complex atomic-level simulations of materials, providing significant insights into the molecular-scale phenomena that govern a wide of range of chemical processes. While such approximations are sufficient at high temperatures and for heavy atoms, classical predictions can deviate substantially from the true behavior described by quantum mechanics. This is distinctly observed when the quantum of vibrational energy ($\hbar\omega$) associated with a relevant mode exceeds its thermal energy ($k_BT$). For systems with light atoms, the quantum regime ($\hbar\omega_{max}>k_BT$) can be well above room temperature, emphasizing a majority of molecular degrees of freedom are sensitive to nuclear quantum effects (NQEs) \cite{Lobaugh1997,Tuckerman1997}. Thus, a quantum level of theory is necessary to describe the behavior of condensed matter with high fidelity and consistent with experimental observations.

% Nuclear quantum effects (NQEs)
Quantum mechanical nuclear delocalization results in zero-point energy (ZPE), tunneling, and quantized energies of the vibrational modes. The resulting energy distribution is frequency-dependent (Bose-Einstein statistics) that differs considerably from the classical equipartition of energy ($1/2k_BT$ per harmonic mode). This directly affects the specific heat ($C_V$), where classical mechanics predicts, for harmonic terms, a temperature-independent value ($3Nk_B$) that is a substantial overestimate in the quantum regime \cite{Berens1983,Hamilton2019}. NQEs have also been observed to greatly influence hydrogen bonding \cite{Li2011}, vibrational spectroscopy \cite{Calvo2014,Hernndez-Rojas2015}, proton-transfer reaction barriers \cite{Hellstrm2018,Ganeshan2022}, Arrhenius kinetics in organic compounds \cite{Datta2008,Gonzalez-James2010}, and even low-temperature plasticity in metals \cite{Proville2012}.

% Relevance of NQEs at extreme conditions
While quantum effects dominate at low temperatures and for light atoms, molecules can still exhibit non-negligible quantum behavior at elevated temperatures. At room temperature, the ZPE of a chemical bond is approximately an order of magnitude higher than its thermal energy \cite{Markland2018}. As an example, the highest vibrational mode of a water molecule (O-H stretch) oscillates at a frequency of $\sim$3450 cm\textsuperscript{$-1$}, equivalent to the thermal energy at $\sim$5000 K \cite{Martnez-Gonzlez2020}. Under high pressure, proton delocalization is pronounced even at high temperatures (750 K) \cite{Ceriotti2014}, which has shown to drastically affect predicted phase transition pressures \cite{Bronstein2014}. Remarkably, under shock loading, classical MD can underestimate the shock temperatures of organic compounds by as much as hundreds of Kelvin \cite{Kress2000,Goldman2009a,Goldman2009b}. This has been exemplified by the inability of MD to reproduce the expected shock-induced chemistry in nitromethane compared with ultrafast spectroscopy \cite{Islam2019,Bowlan2019}.

% Quantum Thermal Bath (QTB)
Current approaches to incorporate NQEs into MD include path integral molecular dynamics (PIMD) and the more approximate quantum thermal bath (QTB). The QTB introduces quantum effects semi-classically by coupling the system to a Langevin thermostat driven by colored noise \cite{Dammak2009}. Specifically, the random force component is filtered such that its power spectral density stems from the quantum fluctuation-dissipation theorem and follows Bose-Einstein statistics \cite{Callen1951}. The vibrational modes are thermally populated based on the average energy of the quantum harmonic oscillator, with an additional kinetic energy contribution to account for ZPE \cite{Barrat2011}. This approach is advantageous as there is negligible increase in the computational cost. For isolated or weakly-anharmonic systems, the QTB accurately produces expected quantum internal energies and vibrational spectra \cite{Calvo2012,Hernndez-Rojas2015}. Under shock loading, QTB-corrected temperature and pressure Hugoniots of methane showed closer agreement with experiments \cite{Qi2012}. This approach has also provided chemical insights, particularly in unimolecular dissociation \cite{Spezia2019} and combustion \cite{Hamilton2019,Qi2013}. However, the QTB loses accuracy in the presence of anharmonicities, where vibrational mode coupling is enhanced \cite{Dammak2011}. This leads to \textit{leakage} of ZPE, where energy is unphysically transferred from high- to low-frequency modes, resulting in an inaccurate quantum energy distribution \cite{Habershon2009}. This has been demonstrated to be a major challenge in predictions of thermal conductivity \cite{Bedoya-Martnez2014} and phase transitions \cite{Hernndez-Rojas2015}. Previous studies have also reported a significant overestimation of kinetic rate constants of unimolecular gas-phase reactions \cite{Angiolari2022} and peak broadening in the vibrational spectra of polyatomic molecules \cite{Calvo2014,Calvo2012}. While the leakage can be reduced with stronger coupling between the thermostat and the system, excessively large damping parameters will cause Langevin dynamics to exhibit undesired Brownian motion \cite{Brieuc2016}.

% Path Integral Molecular Dynamics (PIMD)
Short of solving the Schr\"odinger equation, the quantum behavior of nuclei can alternatively be expressed through the Feynman path integral formalism. With this approach, the probabilistic nature of a quantum particle is captured by considering all possible paths between an initial and final state. Specifically, the quantum mechanical partition function is mapped to a path integral over all trajectories, each weighted by the exponential of its action \cite{Feynman1948}. The resulting Hamiltonian corresponds to an extended classical system of multiple replicas/beads (\textit{P}) of the original system, each subject to a physical potential as well as harmonic interactions with neighboring replicas. This object is referred to as a ring polymer, which more precisely reflects quantum statistical mechanics as it grows in size and becomes exact as $P\to\infty$. Consequently, the computational cost increases, becoming \textit{P} times more expensive relative to a classical simulation. PIMD has been widely employed to comprehensively investigate NQEs in water, particularly the role of ZPE and tunneling on its structural and dynamic properties \cite{Stern2001,Shiga2005,Morrone2008,Habershon2009a,Eltareb2022,Ganeshan2022}. Most notably, the net impact of NQEs on hydrogen bonding is well characterized by the bond strength, which varies depending on the material system. Particularly, with the inclusion of NQEs, weaker H-bonds (e.g. in water) become further weakened, while stronger ones (e.g. in large \ce{HF} clusters) are strengthened. This arises from the ``competition'' of quantum fluctuations affecting intramolecular bond stretching and intermolecular bond bending modes, which respectively strengthen and weaken the H-bond \cite{Li2011}.

\begin{figure}[ht]
    \centering
    \includegraphics[width=1\textwidth]{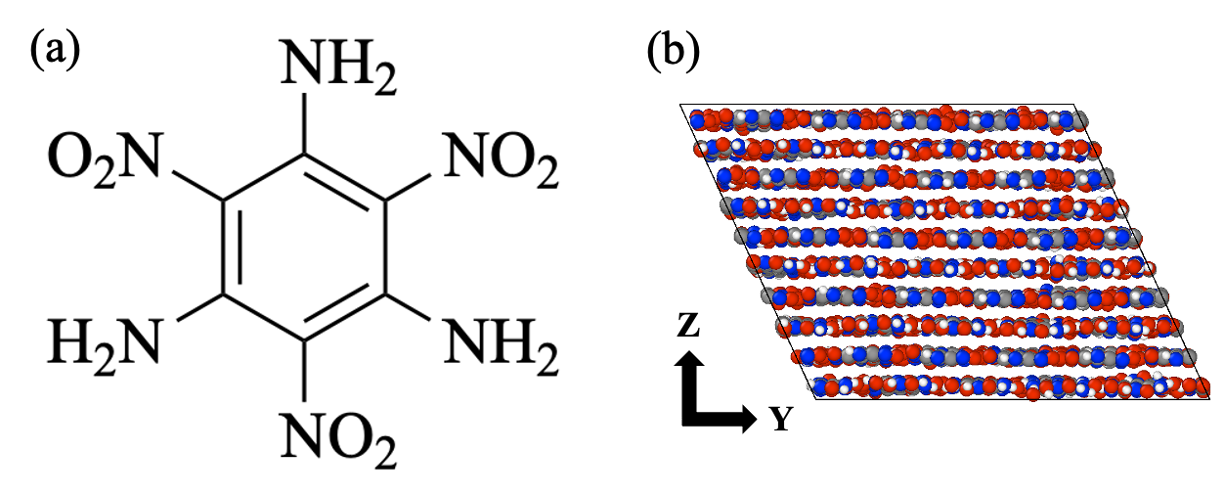}
    \caption{(a) Chemical and (b) crystallographic structure of 1,3,5-triamino-2,4,6-trinitrobenzene (TATB). Atoms are colored as follows: carbon (gray), hydrogen (white), oxygen (red), and nitrogen (blue).}
    \label{fig:molecule}
\end{figure}

% State-of-the-art and Paper Outline
For elementary gas-phase reactions, PIMD has proven to accurately capture chemical kinetics \cite{Collepardo-Guevara2009,Suleimanov2011,PrezdeTudela2012,Li2013}, whereas the QTB overestimates it \cite{Angiolari2022}. However, the application to condensed matter systems with intricate reaction pathways has not yet been thoroughly investigated. To fill this gap, we focus on high explosives (HEs), energetic materials that detonate through rapid, self-sustaining chemical reactions, typically initiated through thermal or shock mechanisms \cite{Lee1980,Tarver2004}. We specifically performed reactive MD on single crystal 1,3,5-triamino-2,4,6-trinitrobenzene, TATB (\ce{C6H6O6N6}), shown in Figure~\ref{fig:molecule}. This is widely known for its insensitivity, characterized by its excellent stability under thermal, mechanical, and shock loading \cite{Rice1990,Dobratz1995}. Its crystal structure, which exhibits near-planar layers of molecules with significant hydrogen bonding networks between the amino and nitro groups \cite{Cady1965}, gives rise to hydrogen transfer dominant reactions \cite{Wu2000,RiadManaa2004}. Prior simulations showed that TATB exhibits increased sensitivity under thermal and shock insults when paired with the QTB \cite{Hamilton2019}. We build upon these findings by conducting simulations with the more quantum-accurate PIMD. Our simulations reveal that PIMD and the QTB lead to noticeable differences in thermal decomposition and reaction kinetics, despite comparable predictions in energies at equilibrium. These are quantified in terms of overall activation barriers, chemical processes throughout all stages of decomposition, and the distribution of energy among the vibrational modes.

% Methods
%%%%%%%%%%%%%%%%%%%%%%%%%%%%%%%%%
\section{Methods}

\subsection{Simulation Details}

We conducted reactive MD simulations using the LAMMPS software package \cite{Thompson2022}, with atomic interactions and chemistry described by the ReaxFF-2018 force field \cite{Wood2018}. ReaxFF potentials have been widely implemented to shed light on atomic-level chemical mechanisms under extreme conditions \cite{Strachan2005a,Wood2015,Islam2019,Islam2020,Kober2022} and support the development of continuum models in HEs \cite{Lee2016,Sakano2020,Lafourcade2023}. Our TATB system consists of 250 molecules and was constructed by replication of the triclinic P$\overline{1}$ unit cell (Figure~\ref{fig:molecule}(b)) \cite{Cady1965}. This was equilibrated at 300 K and 1 atm under isothermal-isobaric (NPT) conditions, resulting in a density of 1.94 g/cm\textsuperscript{3} that aligns closely with experiments \cite{Cady1965}. Isothermal-isochoric (NVT) simulations were then performed across a wide range of temperatures (T = $100-2000$ K) with Langevin-style thermostatting. The thermal state was initialized by assigning a Gaussian distribution of thermal velocities at twice the simulation temperature. This is a standard approach in ReaxFF-MD to minimize undesired reactions during the heating phase, since the kinetic energy quickly convergences to the target value under the condition that the potential energy is initially at equilibrium \cite{Han2011,Kober2022}. All simulations used a 0.1 fs time step and a 100 fs damping parameter. For the QTB, the upper frequency cutoff for the colored-noise thermostat was set to 0.5 fs\textsuperscript{$-1$}, which had previously been determined to produce converged kinetic energy and ZPE predictions for TATB \cite{Hamilton2019}. As for PIMD, the equations of motion were integrated in terms of the normal modes of the ring polymer (\textit{k}), each coupled to its own Langevin thermostat with a corresponding damping parameter calculated with the procedure discussed in Ceriotti \latin{et al.} \cite{Ceriotti2010,Li2026}. The centroid mode (\textit{k} = 0), which corresponds to the spatial average of all replicas, was assigned the same 100 fs damping parameter as the other approaches.

Molecular species were determined using ReaxFF bond orders calculations every 0.1 ps, with a bond order cutoff of 0.5. To investigate how the distribution of energy across the vibrational modes affect chemical processes, vibrational power spectra were obtained through the discrete Fourier transformation of the atomic velocities over a 15 ps NVT simulation (7500 frames) \cite{Berens1983,Wood2014}. Further details on the calculations are provided in Section SM-3 of the Supporting Information. Additionally, we performed a weighted integration of the classical density of states to determine quantum corrections of classical predictions in accordance with Bose-Einstein statistics (qDOS). This has been shown to be extremely precise for perfectly harmonic systems \cite{Berens1983}.   

\subsection{Quantum Convergence and Equilibrium Thermodynamics}

\begin{figure}[ht]
    \centering
    \includegraphics[width=1\textwidth]{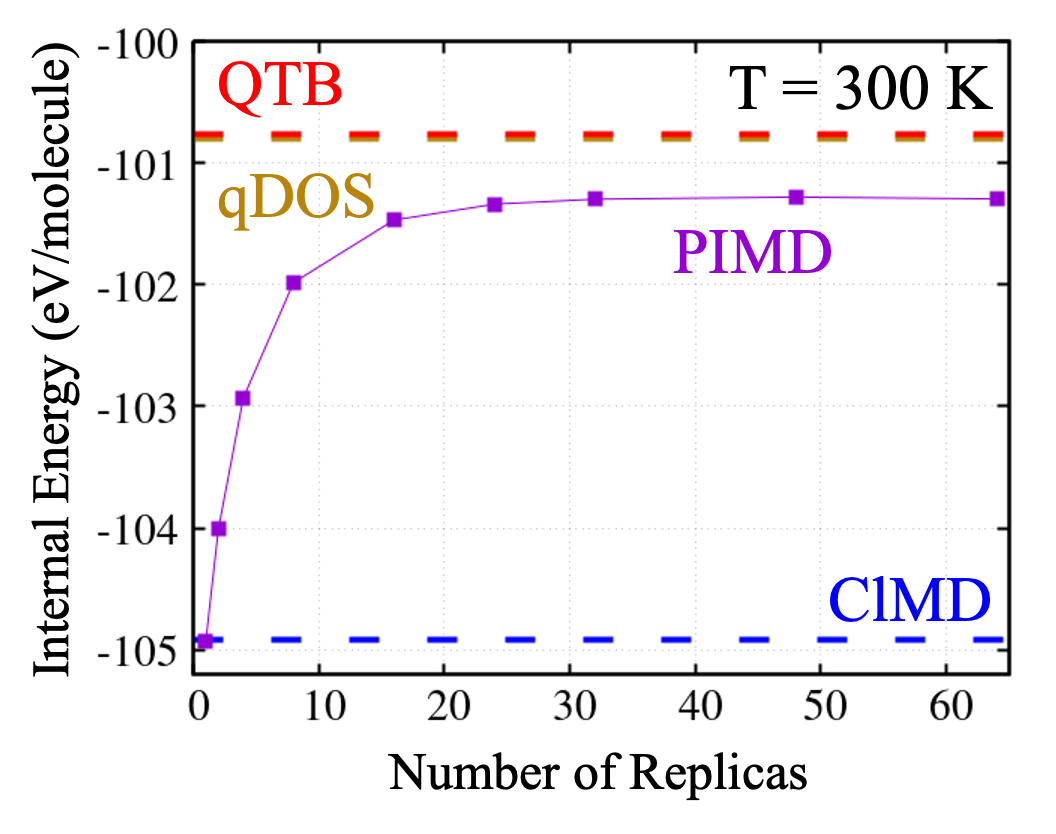}
    \caption{Average molecular energy at 300 K using PIMD as a function of the number of replicas compared with classical (ClMD) and quantum thermal bath (QTB) simulations. Quantum weighted density of states (qDOS) predictions were calculated directly from ClMD trajectories.}
    \label{fig:convergence}
\end{figure}

As the level of quantum representation within PIMD is contingent on the size of the ring polymer, we conducted a convergence study of the internal energy at 300 K. Figure~\ref{fig:convergence} shows the internal energy per TATB molecule, averaged over 15 ps simulations, with respect to the number of replicas \textit{P}. These are compared with classical MD (ClMD), QTB, and qDOS predictions with dashed blue, red, and yellow lines, respectively. As expected, the single-replica ($P=1$) simulation simplifies to the classical Hamiltonian, resulting in an energy almost identical to the ClMD value. While the exact quantum value is achieved for $P=\infty$, convergence is generally reached when \textit{P} exceeds the ratio between the highest quantum of vibrational energy ($\hbar\omega_{max}$) and the thermal energy ($k_BT$) of the system \cite{Markland2018}. We observe convergence in our system roughly at \textit{P} = 24, where the internal energies approach the quantum expectation value. For temperature below this, \textit{P} at convergence is larger due to the stronger influence of NQEs.

\begin{figure}[ht]
    \centering
    \includegraphics[width=1\textwidth]{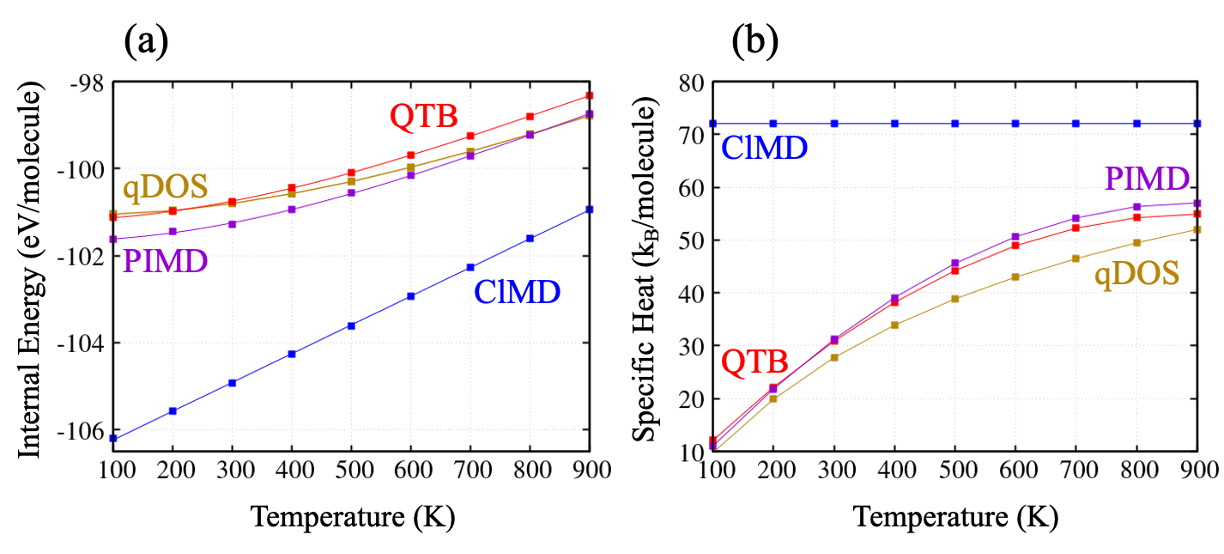}
    \caption{(a) Average molecular energy and (b) specific heat of TATB at various temperatures from PIMD, QTB, and ClMD simulations, as well as qDOS predictions. For temperatures below 300 K, \textit{P} was increased to 128 replicas to accommodate for convergence.}
    \label{fig:energies}
\end{figure}

We then extended these results as a function of temperature up to the reaction thresholds, using $P=32$ and $P=128$ for PIMD simulations above and below 300 K, respectively. The internal energies are shown in Figure~\ref{fig:energies}(a) and compared to ClMD, QTB, and qDOS predictions. Additionally, Figure~\ref{fig:energies}(b) shows the constant-volume specific heat, calculated through numerical differentiation of the energy-temperature relationship in Figure~\ref{fig:energies}(a). Overall, all quantum approaches exhibit the expected temperature-dependent behavior described by quantum theory. The slight overestimation of the QTB internal energies are largely a consequence of pronounced anharmonicity in the molecular crystal, as opposed to isolated molecules \cite{Hernndez-Rojas2015,Brieuc2016}.

% Results and Discussion
%%%%%%%%%%%%%%%%%%%%%%%%%%%%%%%%%
\section{Results and Discussion}

% Overall Chemistry and Kinetics
%%%%%%%%%%%%%%%%%%%%%%%%%%%%%%%%%
\subsection{Thermal Decomposition and Initial Reaction Kinetics}

\begin{figure}[ht]
    \centering
    \includegraphics[width=1\textwidth]{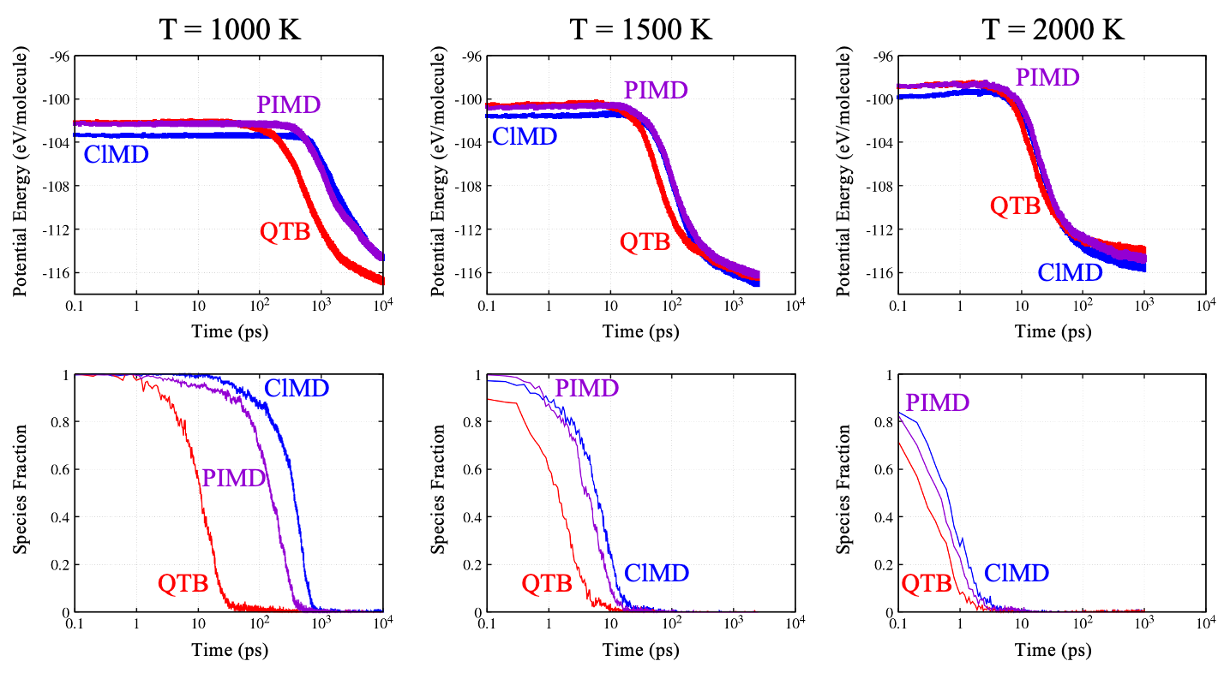}
    \caption{Potential energy (top) and species fraction of TATB (bottom) over time under isothermal-isochoric decomposition at temperatures of 1000, 1500, and 2000 K.}
    \label{fig:pe_decomposition}
\end{figure}

% Trends and comparisons of overall timescales/rates of chemistry
To explore the role of NQEs in the chemical decomposition of crystalline TATB, we performed PIMD ($P=32$), QTB, and ClMD simulations at 1000, 1500, and 2000 K under isothermal and isochoric conditions. Figure~\ref{fig:pe_decomposition} displays the potential energy over time for the various cases (top panels) and the evolution of the fraction of TATB molecules (bottom). The evolution of kinetic energies is shown in Figure S1 of the Supporting Information. The decrease in potential energy aligns with the multi-step formation of exothermic products such as \ce{N2} and \ce{CO2}. At 2000 K, the initial potential energy and its decomposition profile are similar across all simulation methods, as negligible NQEs are expected for such a strongly classical state. As the temperature is reduced, increased zero-point effects shift the initial potential energies of the QTB and PIMD upward with respect to the ClMD case. Notably, even at 1000 K, PIMD predicts a potential energy decrease comparable to classical mechanics. This close agreement indicates that NQEs are of minor significance in the overall exothermicity, consistent with prior simulations of high temperature combustion \cite{Drakon2012,Lele2021}. However, incorporating NQEs via the QTB results in faster reactions and exothermic decomposition for temperatures of 1000 and 1500 K as compared to both PIMD and ClMD.

The evolution of the fraction of TATB molecules indicates that the initial decomposition, dominated by intra- and intermolecular hydrogen transfer from the amino group to the nitro group \cite{Wu2000}, is affected by NQEs, captured by PIMD for the lower two temperatures. The delocalization and tunneling in PIMD result in faster decomposition reactions as compared to the classical model while QTB significantly overestimates this acceleration. Since these initial reactions are only slightly endothermic, the differences between PIMD and ClMD are not manifested in the energy evolution.

\begin{figure}[ht]
    \centering
    \includegraphics[width=1\textwidth]{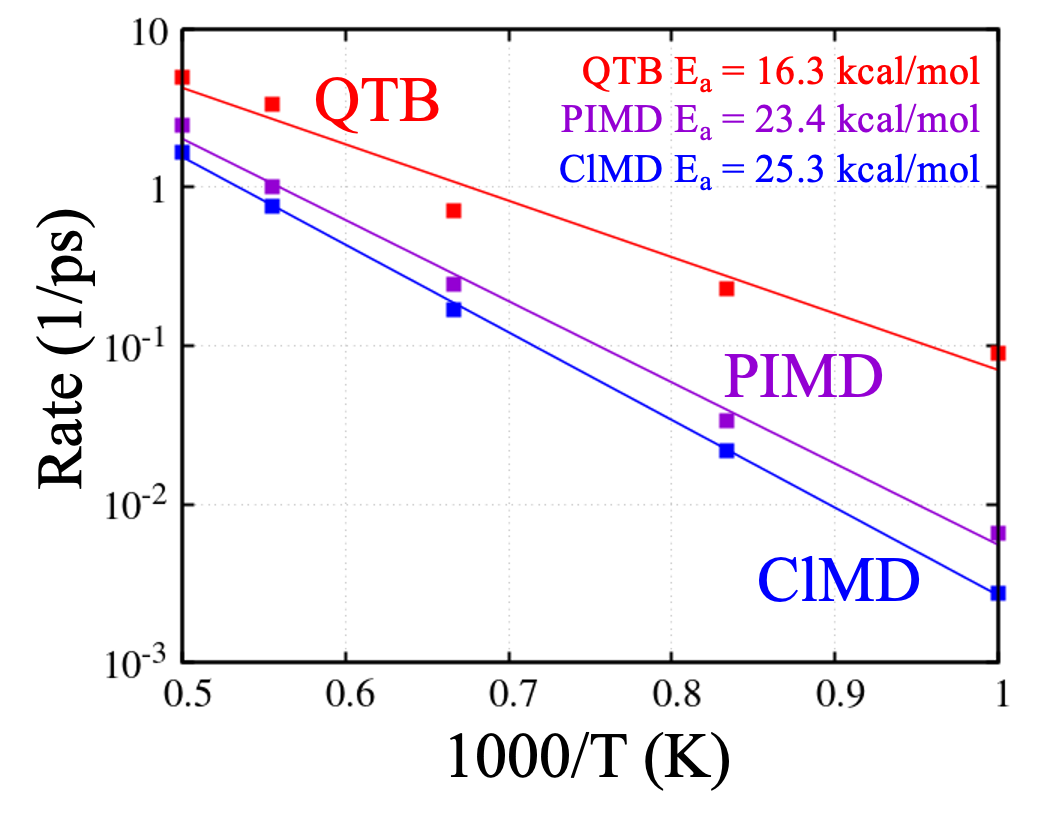}
    \caption{Arrhenius kinetics plots for TATB decomposition with corresponding activation energies ($E_a$). The characteristic time to determine the rates is defined as the point at which 50\% of the TATB molecules remain.}
    \label{fig:kinetics}
\end{figure}

We now quantify differences between the three simulation approaches in describing the overall chemistry by comparing rate constants and activation energies. We define average reaction rates as the inverse of the time corresponding to the reaction of 50\% of the TATB molecules. These rates are shown in Figure~\ref{fig:kinetics} for temperatures ranging from 1000 to 2000 K. As expected, the rates converge towards a unique value with increasing temperature. Similar to what was observed in unimolecular reactions \cite{Angiolari2022}, the higher QTB rates at all temperatures reflect the overly accelerated TATB reaction. The more accurate PIMD method results in rates between the classical and QTB simulations, with NQEs more noticeable at lower temperatures.

Assuming an Arrhenius, single-step chemistry, we obtain average activation energies ($E_a$) by linear fits to the logarithm of the rates as a function of $1/T$. These values are 16.3, 23.4, and 25.3 kcal/mol for QTB, PIMD, and ClMD, respectively. Our simulations indicate that the QTB activation energy is $\sim$43\% lower than that of ClMD, which considerably differs from the $\sim$8\% reduction predicted by PIMD. This is consistent with prior results of unimolecular gas-phase reactions, where the deviation from the classical value was accurately captured with PIMD \cite{Angiolari2022}. As mentioned earlier, the hydrogen-dominant reactions during early-stage decomposition are expected to lower the activation barriers when NQEs are considered. While the PIMD results emphasize that NQEs hold significance in initial reactions, the QTB significantly underestimates such energy thresholds. We discuss the underlying physics behind the QTB and PIMD which lead to these discrepancies in later sections.

% Detailed Chemistry
%%%%%%%%%%%%%%%%%%%%%%%%%%%%%%%%%
\subsection{Multi-step Reaction Pathways}

\begin{figure}[ht]
    \centering
    \includegraphics[width=1\textwidth]{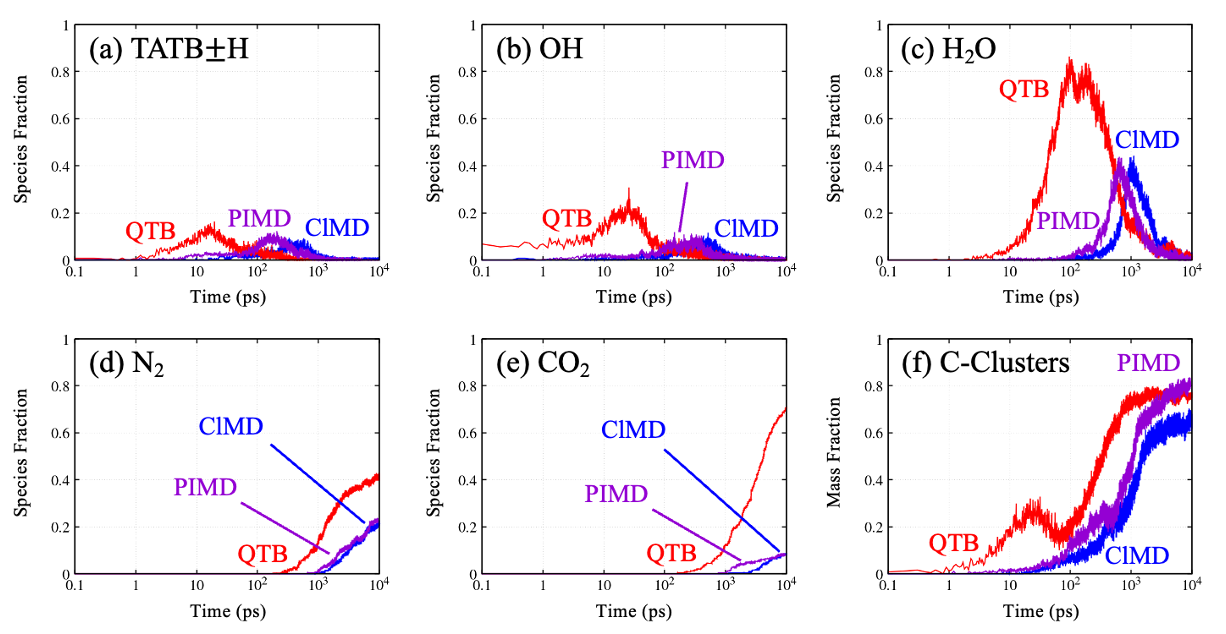}
    \caption{Species time-profiles of (a) TATB$\pm$H (\ce{C6H5O6N6} and \ce{C6H7O6N6}), (b) \ce{OH}, (c) \ce{H2O}, (d) \ce{N2}, (e) \ce{CO2}, and (f) carbon clusters at 1000 K.}
    \label{fig:chemistry}
\end{figure}

% TATB reactions (hydrogen transfer, water formation, detonation products, carbon clusters)
The thermal decomposition of TATB involves a variety of intricate reactions that occur over an extended period of time. To more precisely elucidate the role of NQEs across these reactions, we analyzed the population and mass evolution of key intermediate and product species. This is presented in Figure~\ref{fig:chemistry} for T = 1000 K, normalized by the initial amount of TATB. As previously discussed, the initial reactions are dominated by the intra- and intermolecular transfer of hydrogen to a neighboring oxygen atom. We show this in Figure~\ref{fig:chemistry}(a) in terms of the number of molecules missing a hydrogen atom (\ce{C6H5O6N6}) and with an extra atom (\ce{C6H7O6N6}), denoted as TATB$\pm$H. The small and broad profiles indicate the transient reaction intermediate character of this molecule, as the elimination of \ce{OH} from the nitro group (Figure~\ref{fig:chemistry}(b)) occurs immediately after proton transfer. The earlier onset predicted by PIMD relative to ClMD highlights the degree to which delocalization softens the hydrogen-bond network, therefore enhancing the proton transfer reaction.

The next stage is the formation of \ce{H2O} (Figure~\ref{fig:chemistry}(c)), which occurs through an \ce{OH} elimination followed by a second hydrogen transfer. The nearly concurrent formation of both species, along with low populations of \ce{OH}, indicates that the second hydrogen transfer occurs nearly instantly, in agreement with previous ReaxFF studies \cite{Zhang2009,Bidault2025}. The \ce{OH} evolution in PIMD is only slightly faster than in ClMD. This is expected, as \ce{OH} production is dependent on N-O bond cleavage, which is governed by heavier atoms and, therefore, more classical in nature. The QTB predicts this reaction to begin very early on, peaking at around 10 ps, with a value twice as high compared to the other approaches. This leads to the acceleration of \ce{H2O} formation by nearly an order of magnitude. Additionally, the number of molecules produced at the peak is about double the number observed in the other approaches. Instead, ClMD and PIMD predict a greater proportion of hydrogen atoms remain bound to TATB fragments, which eventually agglomerate into carbon clusters. The sooner than expected initiation of all discussed hydrogen-based reactions suggest that the QTB is overestimating quantum fluctuations, which systemically reduces the reaction time of each intermediate.

The final product stage is marked by the near elimination of \ce{H2O} and the formation of gaseous detonation products, predominantly \ce{N2} (Figure~\ref{fig:chemistry}(d)) and \ce{CO2} (Figure~\ref{fig:chemistry}(e)). The production of \ce{N2} generally follows the \ce{OH} elimination reaction through intermolecular nitroso–imino interactions or, to a lesser extent, between two imino groups \cite{Manaa2010,Tiwari2016}. The highly comparable \ce{N2} profiles between PIMD and ClMD indicate that this reaction is largely classical, which is expected. With respect to \ce{CO2}, its formation is governed by the fragmentation and oxidation of the TATB ring structure. Both PIMD and ClMD predict relatively small and similar populations, attributed to the presence of carbon atoms in high molecular weight clusters. Interestingly, the substantial growth of \ce{CO2} appears a couple of nanoseconds earlier in PIMD than in ClMD. This is primarily due to the difference in timescales associated with \ce{H2O} dissociation, as a large proportion of carbon reacts with the oxygen in \ce{H2O} molecules \cite{He2019}. Once again, the QTB fails to accurately represent the influence of NQEs on these reactions, resulting in faster and increased generation of both products.

Finally, the negative oxygen balance of TATB results in the production of large carbon-rich clusters (soot) \cite{Manaa2009,Zhang2009,Wen2015,Bidault2025}. Carbon clustering follows three primary stages: (1) nucleation of carbon-containing intermediates following initial reactions, (2) aggregation into medium-sized structures and elimination of some non-carbon atoms through gaseous product generation, and (3) coalescence into large carbon-rich foam \cite{Zhang2009}. While the first two stages proceed quickly (ps$-$ns), the final stage can take hundreds of ns \cite{Sollier2022}. We define a cluster as any carbonaceous molecule with mass exceeding that of TATB (258.15 g/mol). The evolution of total mass fraction in carbon clusters is shown in Figure~\ref{fig:chemistry}(f). During the initial phase of decomposition, PIMD and ClMD profiles are similar, suggesting the classical nature of the nucleation process. NQEs become more noticeable in the second phase, shown by marginally higher PIMD values beginning at $\sim$100 ps. With the QTB, clustering begins in parallel with hydrogen-transfer reactions and exhibits a modest decrease coinciding with \ce{H2O} formation, before steadily increasing again. The QTB pathway unfolds over an order magnitude faster than PIMD and ClMD. Interestingly, QTB and PIMD reach similar mass fractions in carbon clustering within the simulation timescale ($\sim$76\% and $\sim$80\%, respectively) whereas the value for ClMD is slightly lower ($\sim$65\%). By analyzing the cluster populations, we find that the QTB total mass is concentrated into 4 clusters (average mass of 12231 g/mol), markedly fewer in number than ClMD (27 clusters with average mass of 1558 g/mol) and PIMD (9 clusters with average mass of 5747 g/mol). Thus, although the final stage of clustering takes places over timescales beyond our simulations, PIMD predicts this process to proceed more rapidly, whereas the QTB exaggerates the acceleration. We note that these results do not reflect long-term chemical equilibrium, which necessitate larger system sizes and extended simulation times.

At higher temperatures (1500 and 2000 K), all reactions initiate sooner. The hydrogen-based reactions begin almost immediately and behave classically. Formation of \ce{N2} remains similar across all approaches, but decreases due to the greater populations of \ce{NH3}. While the total cluster mass is similar at 1500 K, it decreases at 2000 K due to the significant \ce{CO2} production. Surprisingly, \ce{CO2} populations at 1500 K are noticeably disparate, with PIMD yielding the lowest amount. This could imply that carbon remains attached to clusters during the last stage of agglomeration, which is not captured within our simulation timescales at 1000 K. We show these results in Figures S2 and S3 of the Supporting Information, respectively.

% PIMD vs. QTB
%%%%%%%%%%%%%%%%%%%%%%%%%%%%%%%%%
\subsection{Origin of the Discrepancy Between PIMD and QTB}

% Physics-based differences between QTB and PIMD that explains the above results
To gain a deeper understanding of the observed differences in chemistry between the three methods, we analyzed the key physics-based distinctions between the QTB and PIMD approaches. In PIMD, the quantum partition function is sampled via the dynamics of a classical ring polymer and physical observables are determined through estimators, which connect the ring polymer representation to quantum mechanical expectation values. Estimators of observables that depend only on position (e.g. potential energy and bond order) are simply an average over all replicas \cite{Tuckerman2023}. However, the kinetic energy operator ($\hat{T} = -\frac{\hbar^2}{2m} \nabla^2$) is embedded in the harmonic spring terms of the Hamiltonian that include couplings between replicas. The most efficient estimator is the centroid-virial estimator, derived from the virial theorem using integration by parts of the kinetic energy expectation value \cite{Ceperley1995}:

\begin{equation}
    KE = \frac{3}{2} N k_B T - \frac{1}{2P} 
    \sum_{j=1}^{P} \sum_{i=1}^{N}
    \left( q_i^{[j]} - \bar{q}_i \right)
    \frac{\partial V\!\left( q_1^{[j]}, \ldots, q_N^{[j]} \right)}{\partial q_i^{[j]}}
    \label{eqn:cv_estimator}
\end{equation}

where $q_i^{[j]}$ denotes the position of replica \textit{j} of atom \textit{i} and $\bar{q}_i$ is the position of the corresponding ring polymer centroid. The first term corresponds to the classical energy, while the second term incorporates quantum corrections arising from the spread of the replicas. The ring polymer's collective motion is represented by the centroid, enabling a direct comparison of its real-time chemical evolution with the other approaches.

\begin{figure}[ht]
    \centering
    \includegraphics[width=1\textwidth]{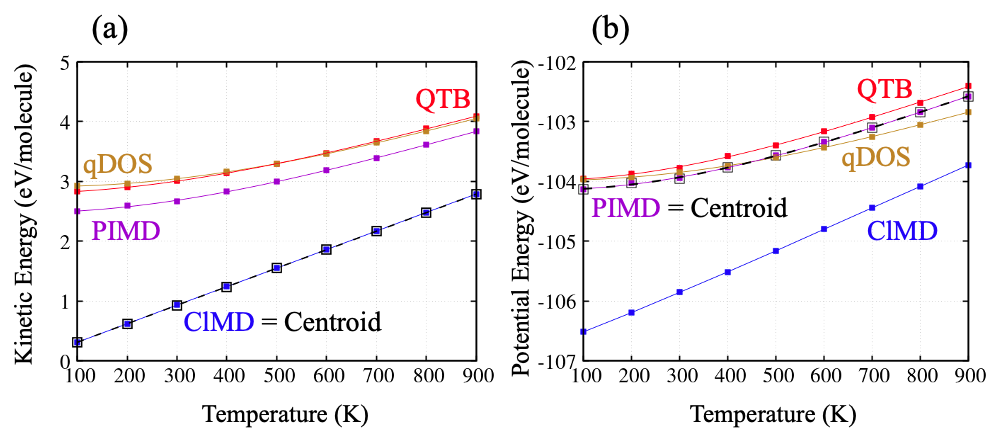}
    \caption{Average molecular (a) kinetic and (b) potential energy of TATB at various temperatures from PIMD, QTB, and ClMD simulations. These are additionally compared with qDOS predictions and energies of the ring polymer centroids. For temperatures below 300 K, \textit{P} was increased to 128 replicas to accommodate for convergence. The black curves indicate that ClMD and centroid in (a), and PIMD and centroid in (b), are effectively identical.}
    \label{fig:centroid_compare}
\end{figure}

As for the QTB, quantum effects are introduced by adding frequency-dependent (colored) noise to classical dynamics. Similar to ClMD, expectation values are then computed as ensemble averages over the atomic trajectories. To highlight the limitations of QTB for computational chemistry, we revisit the comparisons of internal energies in Figure~\ref{fig:energies}(a), but this time separating the various contributions, as seen in Figure~\ref{fig:centroid_compare}. In addition to potential and kinetic energies, we include the kinetic energy of the PIMD centroids, calculated from the zeroth mode of the ring polymers. As with the internal energy, the individual energy components in PIMD and QTB are in good agreement. However, the centroid kinetic energy in PIMD is significantly lower that of the QTB value; this overestimation of the velocities in the QTB results in unrealistically accelerated chemistry. This difference can be understood based on how each approach treats ZPE and quantum fluctuations. In PIMD, ZPE is distributed among the internal modes of the ring polymer rather than the centroid, as reflected in the second term of Eq.~\ref{eqn:cv_estimator}. Thus, while the centroid motion is classical (Figure~\ref{fig:centroid_compare}(a)), it is influenced by a quantum potential generated by the fluctuations of the higher order modes (Figure~\ref{fig:centroid_compare}(b)) \cite{Ceriotti2010}. In contrast, while the QTB roughly produces the expected quantum potential energy, ZPE is assigned directly to the physical particle as kinetic energy, which leads to an effective temperature considerably larger than the simulation temperature. Thus, while the QTB can accurately produce energy expectation values at equilibrium, it artificially accelerates chemical reactions and significantly underestimates activation barriers.

We further illustrate the limitations of the kinetic energy formulation in the QTB in the context of vibrational spectral features, which provide valuable insights on chemical bonding, molecular interactions, and reaction pathways at ultrafast timescales \cite{Mcgrane2009,Islam2019,Powell2020}. Regardless of its associated vibrational frequency, classical mechanics distributes an average energy of $1/2k_BT$ to each quadratic degree of freedom. In contrast, the QTB partitions energy according to the Bose-Einstein distribution, with an additional term accounting for the ZPE. This gives rise to higher frequency modes gaining energy and lower frequency modes losing energy relative to a classical system. For PIMD, the quantum mechanical energy distribution is more accurately produced through Feynman's path integral formulation at convergence. 

\begin{figure}[ht]
    \centering
    \includegraphics[width=1\textwidth]{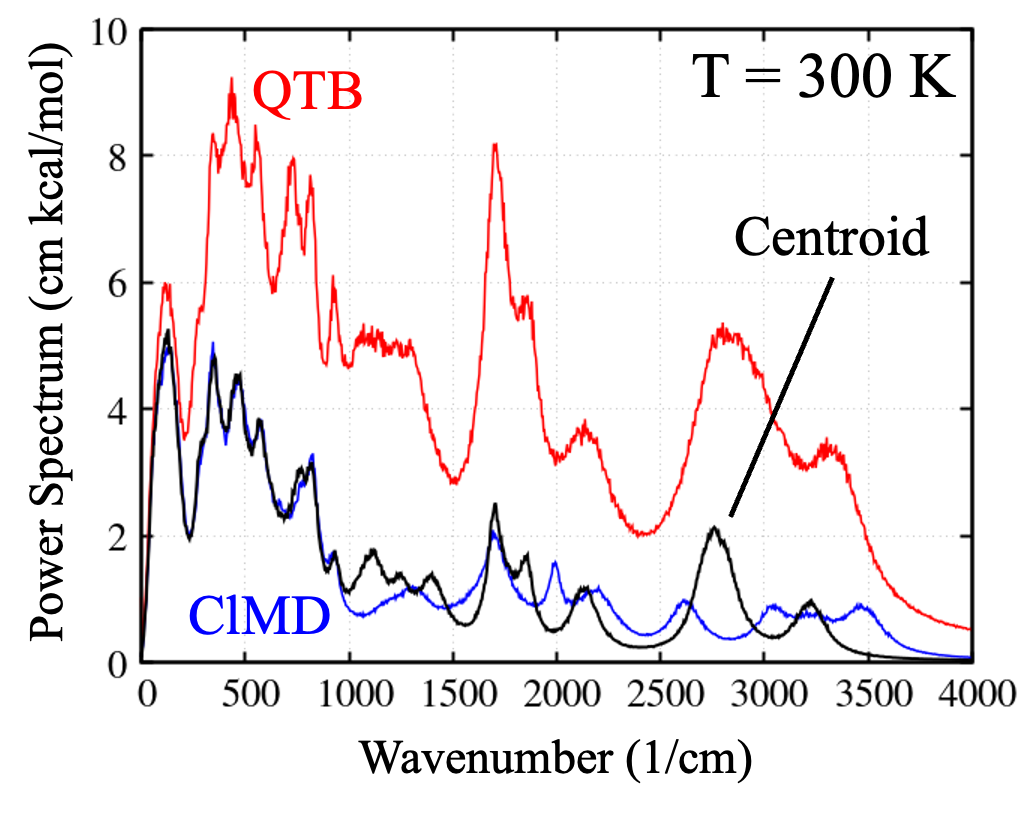}
    \caption{Vibrational power spectra at 300 K. The centroid spectrum corresponds to the zeroth-order normal modes of the ring polymers in PIMD.}
    \label{fig:powerspec}
\end{figure}

Power spectra from the 300 K simulations are shown in Figure~\ref{fig:powerspec}. Integration of these curves over frequency yields the kinetic energy value present in Figure~\ref{fig:centroid_compare}(a) for the respective temperature. The low-frequency modes (ring twists, distortions, and lattice vibrations) are largely classical, as their thermal energy is comparable to the ZPE ($\hbar\omega/2$). Thus, the centroid is nearly identical to the classical spectrum, with the QTB showing a similar profile in this regime. For the highest frequency modes, corresponding to \ce{NH2} stretching in TATB \cite{Liu2006a}, quantum fluctuations and ZPE cause slight peak broadening and redshifting \cite{Calvo2014}. While we observe such behavior in the QTB and centroid curves, the QTB exhibits a noticeably higher spectrum across all modes due to its incorporation of NQEs via the kinetic energy, which inadvertently leads to additional peak softening. We found that this had significant consequences on the QTB predictions of chemical initiation and thermal decomposition rates.

Even though the QTB assigns significantly more energy to every mode, the peak positions, notably for modes exceeding 1500 cm\textsuperscript{$-1$}, are comparable with the centroid spectrum. Slight discrepancies are seen with the highest frequency peaks, where the centroid spectrum is redshifted relative to the QTB. The centroid peak intensities are similar in magnitude to those of the classical spectrum, reflecting how ClMD and PIMD produce comparable results during thermal decomposition. At temperatures closer to the onset of reactivity, the power spectra are in closer agreement, yet the QTB remains marginally elevated. We show this at 800 K in Figure S4 of the Supporting Information.

% Conclusions
%%%%%%%%%%%%%%%%%%%%%%%%%%%%%%%%%
\section{Conclusions}

NQEs in condensed matter systems cannot be ignored because of the notable influence of zero-point motion, delocalization, and tunneling on structural, thermodynamic, and chemical behavior. PIMD is a well-established approach to naturally incorporate NQEs by representing a quantum particle as a classical ring polymer of many replicas. We used this approach, along with the more approximate quantum thermal bath, to investigate how NQEs influence the multi-step reactions and kinetics of the thermal decomposition of the energetic material TATB. Although both approaches yield the expected quantum internal energies at equilibrium, they differ with respect to the incorporation of quantum effects. Specifically, the QTB approximates NQEs as an additional energy contribution through the momenta of the physical particles. As a consequence, the kinetic energies at a given temperature are substantially different between classical and QTB simulations. We find that, under constant temperature, this overestimates delocalization, inducing an artificial acceleration of chemical reactions and an overly pronounced reduction of initial reaction barriers. Analysis of the vibrational spectra also revealed an excessive broadening of the high-frequency peaks.

In contrast, the more accurate PIMD accounts for zero-point energy via the internal vibrational modes of the ring polymer. The centroid, which represents the average position of the ring polymer's replicas, evolves according to classical mechanics but under forces that encompass quantum fluctuations. While the PIMD decomposition profiles and activation energies align more closely with those of ClMD, certain reactions vary, especially at 1000 K. The influence of proton delocalization on hydrogen transfer and \ce{OH} elimination are more precisely considered, leading to slightly earlier formation of \ce{H2O} and \ce{CO2}. Other reactions, such as \ce{N2} formation and the initial stages of carbon clustering, were observed to behave primarily classically. At higher temperatures, NQEs are diminished, and all approaches predictably converge to classical behavior. Overall, these simulations emphasize the importance of NQEs in the atomistic modeling of not only energetic materials, but in condensed matter involving complex multi-step reactions, while also highlighting the limitations of semi-classical approximations. Our findings motivate additional simulations over an extended range of thermodynamic states to further improve mesoscale reaction chemistry models.

% Acknowledgments
%%%%%%%%%%%%%%%%%%%%%%%%%%%%%%%%%
\begin{acknowledgement}
    The authors thank Brenden Hamilton for providing insights on the Arrhenius kinetics analysis of TATB decomposition and Edward Kober for valuable discussions about nuclear quantum effects. This research was sponsored by the Army Research Office and was accomplished under Cooperative Agreement No. W911NF-22-2-0170. The views and conclusions contained in this document are those of the authors and should not be interpreted as representing the official policies, either expressed or implied, of the Army Research Office or the U.S. Government. The U.S. Government is authorized to reproduce and distribute reprints for Government purposes notwithstanding any copyright notation herein. The authors acknowledge computational resources from nanoHUB and Purdue University through the Network for Computational Nanotechnology and the Los Alamos National Laboratory (LANL) Institutional Computing Program.
\end{acknowledgement}

% Supporting information
%%%%%%%%%%%%%%%%%%%%%%%%%%%%%%%%%
\begin{suppinfo}
    Kinetic energy decomposition profiles, chemical species profiles at elevated temperatures, and vibrational power spectra at 800 K.
\end{suppinfo}

% References
%%%%%%%%%%%%%%%%%%%%%%%%%%%%%%%%%
\bibliography{library}
%\printbibliography

% TOC Graphic
%%%%%%%%%%%%%%%%%%%%%%%%%%%%%%%%%
%\begin{tocentry}
%    \includegraphics[width=\linewidth,height=1.75in,keepaspectratio]{TOC.png}
%\end{tocentry}

\end{document}